# Spectral Phase Effects in High-Order Above Threshold Ionization of Noble Gas Atoms

A. L. Harris*

Physics Department, Illinois State University, Normal, IL, USA 61790

**Abstract**

We present theoretical studies of above threshold ionization (ATI) using sculpted laser pulses for noble gas atoms. The time-dependent Schrödinger equation is solved to calculate the ATI energy and momentum spectra, and a qualitative understanding of the electron motion after ionization is explored using a classical model that solves Newton's equation of motion. Results are presented for Gaussian and Airy laser pulses with identical power spectra, but differing spectral phases. The simulations show that the third order spectral phase of the Airy pulse, which can alter the temporal envelope of the electric field, causes changes to the timing of ionization and the dynamics of the rescattering process. Specifically, the use of Airy pulses in the ATI process results in a shift of the Keldysh plateau cutoff to lower energy due to a decreased pondermotive energy of the electron in the laser field. Additionally, the side lobes of the Airy laser pulse change the number and timing of rescattering events, which results in changes to the high-order ATI plateau. Our results also show that laser pulses with identical carrier envelope phases and nearly identical envelopes yield different photoelectron momentum densities, which are a direct result of the pulse's spectral phase.

## I. Introduction

One of the most instructive and valuable processes in attosecond and strong-field physics is that of above threshold ionization (ATI), in which an atom absorbs more photons than are required for ionization. The excess absorbed energy is converted to the kinetic energy of the ionized electron and the corresponding ATI spectrum has proven to be a valuable tool in many applications. In general, the ATI process can be largely understood using the three-step recollision model in which an electron is ionized and 'born' into the continuum (step 1), after which it accelerates in the electric field. When the electric field changes sign, it causes the electron to reverse direction (step 2) and recollide with the parent ion (step 3). If the electron is elastically scattered from the ion, it can then be accelerated again by the electric field before making its way to the detector. This leads to the so-called high-order ATI (HATI) electrons, which have been shown to be of use in reconstructing electron-ion scattering cross sections that contain target structure information [1–4]. HATI electrons are also used in the laser induced electron diffraction technique (LIED), in which the recolliding electrons are used to probe the target structure with high temporal and spatial resolution [5]. Additionally, the ATI process can be used to characterize the carrier envelope phase of the laser pulse, a quantity that is vital to accurately understanding processes involving few-cycle pulses [6].

Since the ATI process was first observed [7], there have been countless experimental and theoretical studies aimed at elucidating its dynamics and developing applications based on the process. Despite these decades of study, the ATI process still has insights to share. In this work, we combine the familiar ATI process with the use of sculpted laser fields [8–11]. Unlike their more traditional sin-squared or Gaussian wave forms, sculpted wave forms can have more complicated envelope functions with multiple peaks, carry quantized orbital angular momentum, or exhibit self-acceleration, self-healing, and limited diffraction [12–14]. There are several studies using sculpted wave forms to examine processes such as photoionization [15–17], atomic and molecular electronic and rotational transitions [18–22], high order harmonic generation [23–26], and strong field physics with twisted electrons [27,28]. These studies have demonstrated novel physics, including alteration of selection rules [16,22], orbital angular momentum transfer [18,19], the production of twisted UV light [23–26], and the ability to control the orbital angular momentum of XUV probes [23,25,29]. However, the ATI process, and the effects of sculpted wave packets on the ATI energy and momentum spectra, have received only limited attention for spatially sculpted wave packets [30–32]. In these studies, it was demonstrated that a vortex pulse alters the ATI energy spectrum with an increased ionization probability as the orbital angular momentum of the vortex increases. Additionally, the photoelectron is accelerated in the laser propagation direction, something that does not occur for plane wave pulses.

* alharri@ilstu.edu

Here, we present a theoretical study of ATI in noble gas atoms using few-cycle *temporal* Airy and Gaussian laser pulses. These pulses are carefully chosen such that their power spectra are identical, but they have different spectral phases. We aim to determine whether the spectral phase of the Airy pulse alters the photoelectron momentum and if this phase is imprinted on the photoelectron wave packet. If the temporally sculpted laser pulse alters the photoelectron wave packet, then the rescattering dynamics that lead to ATI and other higher order processes are also expected to change. The introduction of the Airy spectral phase as another control parameter in the ATI process could lead to new opportunities for examining target structures through LIED or generating higher order sculpted pulses via HHG.

We present numerical simulations of ATI for hydrogen, helium, neon, argon, xenon, and krypton using the time-dependent Schrödinger equation (TDSE) and combine this with analysis from a classical model that is aimed at elucidating the effects of the sculpted laser fields on the ionized electron dynamics. We show that the direct and rescattering plateau cutoffs in the ATI energy spectrum shift in energy for Airy pulses, and our classical simulations indicate that this shift can be traced to altered pondermotive energy and electron trajectories in the laser field following ionization. Additionally, we show that laser pulses with identical carrier envelope phases and nearly identical envelopes yield different photoelectron momentum densities, which are a direct result of the pulse spectral phase.

The remainder of the paper is organized as follows. Section II contains the basic theoretical and numerical methods used. Section III presents the results along with a discussion of their significance. Lastly, section IV provides a brief summary and outlook. Atomic units are used throughout unless otherwise noted.

## II. Theory
### A. TDSE

Because the primary dynamics of ATI with linearly polarized pulses happen along the laser polarization direction, a one-dimensional approximation is typically sufficient to capture the important physics. We solve the one-dimensional TDSE for a single active electron atom in a laser field

$$\frac{i\partial}{\partial t}\psi(x,t) = \left[-\frac{1}{2}\frac{d^2}{dx^2} + V_a(x) + xE(t)\right]\psi(x,t). \tag{1}$$

The atomic potential $V_a(x)$ is approximated with the pliant core model [33]

$$V_a(x) = -\frac{1}{(|x|^\alpha + \beta)^{1/\alpha}}, \tag{2}$$

which has been shown to most reasonably approximate the results of a full three-dimensional TDSE calculation for HHG [33] and ATI [34]. For the pliant core model, $\alpha = 1.5$ and the values for $\beta$ are given in Table 1 for the different noble gas atoms.

| Atom | $\beta$ for pliant core potential of Eq. (2) |
|---|---|
| H | 1.45 |
| He | 0.49 |
| Ne | 0.62 |
| Ar | 1.11 |
| Kr | 1.37 |
| Xe | 1.78 |

Table 1 Parameter $\beta$ for the pliant core potential function for the noble gas atoms [33].

Two different temporal laser pulse fields $E(t)$ are used: a Gaussian pulse and a truncated Airy pulse. The untruncated Airy function is a solution to the free particle Schrödinger equation [14] with infinite transverse extent and infinite energy (much like the plane wave). Alternatively, the truncated Airy pulse is a finite-width wave packet with finite energy and is a more reasonable model of a physical Airy laser pulse. Additionally, the truncated Airy pulse provides a tunable parameter $\phi_3$ that can be used to adjust the shape of the envelope (see Fig. 1). The Gaussian pulse is given by

$$E_G(t) = E_0 e^{-2\ln 2 \left(\frac{t-t_c}{\Delta t}\right)^2} \sin(\omega_0(t - t_c)) \tag{3}$$

and the truncated Airy pulse is given by [35]

$$E_A(t) = E_0 \sqrt{\frac{\pi}{2\ln 2} \frac{\Delta t}{\tau_0}} Ai\left(\frac{\tau - (t-t_c)}{\Delta \tau}\right) e^{\frac{\ln 2\left(\frac{2\tau}{3} - (t-t_c)\right)}{2\tau_{1/2}}} \sin(\omega_0(t - t_c)), \tag{4}$$

where $\tau_{1/2}$ is the exponential truncation half-life of the Airy, $\omega_0$ is the carrier frequency, $\Delta \tau$ is the stretch of the Airy, $\Delta t$ is the FWHM of the temporal intensity, $\tau$ is the shift of the Airy, $t_c$ is the center of the pulse, and $\tau_0 = \left(\frac{|\phi_3|}{2}\right)^{1/3}$ is a parameter related to the third order term $\phi_3$ of the spectral phase. The third order spectral phase controls the temporal envelope of the field and is related to the truncation half-life, stretch, and shift of the Airy by

$$\tau_{1/2} = \frac{2(\ln 2)^2 \phi_3}{\Delta t^2} \tag{5}$$

$$\Delta \tau = \tau_0 \text{sign}(\phi_3) \tag{6}$$

$$\tau = \frac{\Delta t^4}{32(\ln 2)^2 \phi_3}. \tag{7}$$

Figure 1 depicts the Airy and Gaussian pulses used here. As the magnitude of the third order spectral phase decreases, the Airy pulse envelope becomes more Gaussian-like and its width decreases. The number of side lobes of the Airy pulse also decreases with decreasing magnitude of the third order spectral phase. If $\phi_3$ is negative, the orientation of the Airy pulse is reflected about its center. We refer to these pulses as inverted Airy pulses. It should be noted that the Gaussian, Airy, and inverted Airy pulses are identical with respect to carrier envelope phase. Thus, the zeros of the electric field occur at the same instances in time, however the maximum and minimum of the electric field may occur at different instances and with different magnitudes based on the pulse envelope.

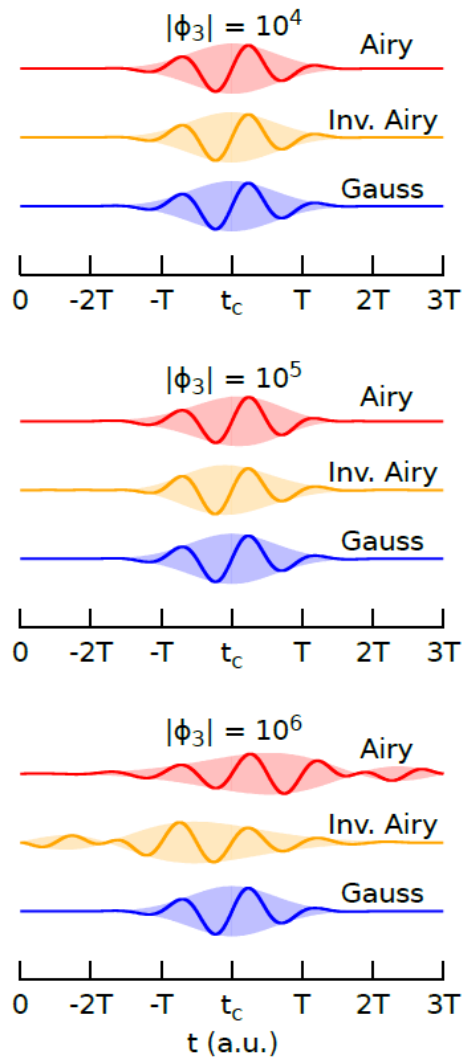

Figure 1 Temporal electric fields for the 6-cyle Airy and Gaussian laser pulses used in the calculations. The shaded areas represent the pulse envelope. Pulse parameters were $\omega_0 = 0.057$

a.u. ($\lambda = 800$ nm), $E_0 = 0.0533$ a.u. ($I = 1 \times 10^{14}$ W/cm$^2$), $\Delta t = 110$ a.u., $t_c = 330.5$ a.u. The third order phase terms $\phi_3$ for the Airy and inverted Airy pulses are shown in the figure.

A unique feature of the Gaussian and Airy pulses used here is that their spectral intensities are identical, and the frequency domain electric fields differ only by a phase. The Gaussian frequency domain electric field is

$$E_G(\omega) = \frac{E_0 \Delta t \sqrt{\pi}}{2^{3/2} \sqrt{\ln 2}} e^{-\frac{\Delta t^2 (\omega_0 - \omega)^2}{(8 \ln 2)}} \tag{8}$$

and the Airy frequency domain electric field is [35]

$$E_A(\omega) = E_G(\omega) e^{-\frac{i}{6}\phi_3 (\omega - \omega_0)^3}, \tag{9}$$

where $\phi_3$ is the same third order term of the spectral phase as described above.

The pulses used here were truncated to include 6 cycles of the electric field. For values of $|\phi_3| \leq 10^5$, this truncation had no effect of the spectral intensity and the Gaussian, Airy, and inverted Airy pulses. However, for $|\phi_3| > 10^5$, the Airy and inverted Airy spectral intensities were altered as a result of the truncation and were no longer identical to that of the Gaussian pulse.

In solving the TDSE, the initial state atomic wave function was found by imaginary time propagation, and the 1D TDSE was solved using the Crank-Nicolson method [36]. Absorbing boundary conditions [37] were used to prevent reflections from the grid boundary and the density of the wave function was checked at each time step to ensure that no probability was lost. The ATI energy spectrum was calculated using the window operator technique [38,39]. All codes are available through figshare.com [40–42] and details of the numerical calculations are provided in Appendix A.

### B. Classical Model

Once the electron has been ionized from the atom, it moves in the presence of a classical electric field. Thus, it is possible to calculate its classical trajectory and kinetic energy at the detector by solving Newton's equation of motion

$$\ddot{x} = -E(t). \tag{10}$$

We modified the ClassSTRONG program [43] to solve Eq. (10) for the ionized electron motion in either an Airy, inverted Airy, or Gaussian laser pulse. From this, the kinetic energies of the direct and rescattered electrons were found as a function of ionization time. Additionally, the times of recollision and the classical trajectories were calculated from solving Eq. (10). Results from the classical models are used to provide qualitative physical insight into the electron dynamics for the different pulse types.

### III. Results and Discussion
### A. Hydrogen Energy Spectra

In general, the ATI energy spectrum shows two distinctive plateaus that occur in regions of low and high energy. In the Keldysh regime [44], photoelectron kinetic energies are below approximately $2U_p$, where $U_p = \frac{e^2 A_0^2}{4m}$ is the pondermotive energy [45]. In this regime, the ATI electrons are a result of direct ionization in which the electrons make their way to the detector without further interaction with the parent ion. The second plateau that is observed in the ATI spectrum is referred to as the high-order ATI (HATI) plateau, and it occurs for photoelectrons with energies between about $2U_p$ and $10U_p$. The HATI plateau can be largely understood through the latter two steps of the three-step recollision model. Following ionization, the electron is accelerated by the electric field, which ultimately changes sign, causing the electron to reverse direction (step 2) and recollide with the parent ion (step 3). If the electron is elastically scattered from the ion, it can then be accelerated again by the electric field before making its way to the detector. The maximum kinetic energy of the rescattered electron occurs for the backscattering geometry, and classical momentum conservation yields a cutoff energy of approximately $10U_p$ [46]. The probability of HATI electrons is approximately the same for energies between 2 and $10U_p$, leading to the HATI plateau.

We performed calculations for an 800 nm laser pulse ($\omega_0 = 0.057$ a.u.) with intensity of $I = 1 \times 10^{14}$ W/cm² ($E_0 = 0.0533$ a.u.) and duration of 6 cycles. The pulse had either a Gaussian or truncated Airy envelope with full-width half-max of $\Delta t = 110$ a.u. temporally centered at $t_c = 330.5$ a.u. Results are presented for Airy pulses with third order phase terms $10^4 \leq |\phi_3| \leq 10^6$. We begin with a detailed discussion of the ATI spectra for hydrogen atoms, and in section III.C, expand to examine spectra for all noble gas atoms.

Figure 2 shows the ATI energy spectra for a model hydrogen atom for Gaussian, Airy, and inverted Airy pulses. For the Gaussian pulse, the Keldysh and HATI plateaus are present, but shifted from their usual $2U_p$ and $10U_p$ cutoff energies to $3U_p$ and $8U_p$. This is due to the small $\Delta t$. For the Airy and inverted Airy pulses with $|\phi_3| = 10^4$, the spectra are similar to that produced by the Gaussian pulse, which is expected since the electric fields are similar (see Fig. 1). As $|\phi_3|$ increases, the low energy ATI spectrum cutoff decreases to around $1.5U_p$, indicating that the direct electrons liberated by Airy or inverted Airy pulses gain less kinetic energy during their time in the laser field than those liberated by Gaussian pulses. This shift in low-energy cutoff can be traced to a decreased maximum electric field of the Airy and inverted Airy pulses relative to the Gaussian pulses. For example, for $|\phi_3| = 10^6$, the maximum electric field strength of the Airy pulse is 83% of the maximum for the Gaussian pulse. The reduced magnitude of the Airy or inverted Airy electric field causes a reduction in the pondermotive energy, which shifts the low-energy cutoff to smaller kinetic energies. Larger values of $|\phi_3|$ result in smaller values of the pondermotive energy and the cutoff shifts to even lower energy. These results are confirmed by the classical calculations shown below in section III.B. The values of the pondermotive energies relative to the Gaussian pulse pondermotive energy are listed in Table 2 for the different Airy and inverted Airy pulses used here.

| Pulse type | $U_p$ relative to Gaussian $U_p$ |
| --- | --- |
| Gaussian | 1 |
| Airy/Inverted Airy $|\phi_3| = 10^4$ | 1 |
| Airy/Inverted Airy $|\phi_3| = 10^5$ | 0.98 |
| Airy/Inverted Airy $|\phi_3| = 10^6$ | 0.68 |

Table 2 Pondermotive energies relative to the Gaussian value for the Airy and inverted Airy pulses used in the calculations.

The HATI plateau is observable in the Gaussian pulse ATI spectrum from approximately 6 to $8U_p$. For the Airy pulses, this plateau shifts in energy and magnitude depending on the value of $\phi_3$. For $\phi_3 > 0$, the HATI plateau shifts to larger energies when $\phi_3 = 10^5$, and maintains its magnitude. For $\phi_3 = 10^6$, the HATI plateau shifts to lower energies and decreases in magnitude. For the inverted Airy pulse, the HATI plateau shifts to much larger energies at $\phi_3 = -10^5$ and decreases significantly in magnitude. However, for $\phi_3 = -10^6$, the HATI plateau resembles that of the Airy pulse, occurring at lower energies and with a reduced magnitude.

The physical explanation of these shifts is somewhat complicated due to the rescattering dynamics involved. Certainly, a reduction in the pondermotive energy for electrons moving in the Airy or inverted Airy fields can cause a shift in the plateau cutoff to lower energy. However, as we detail below in section III.B, the number and timing of the rescatterings changes for the different pulse types and this affects the ATI spectra.

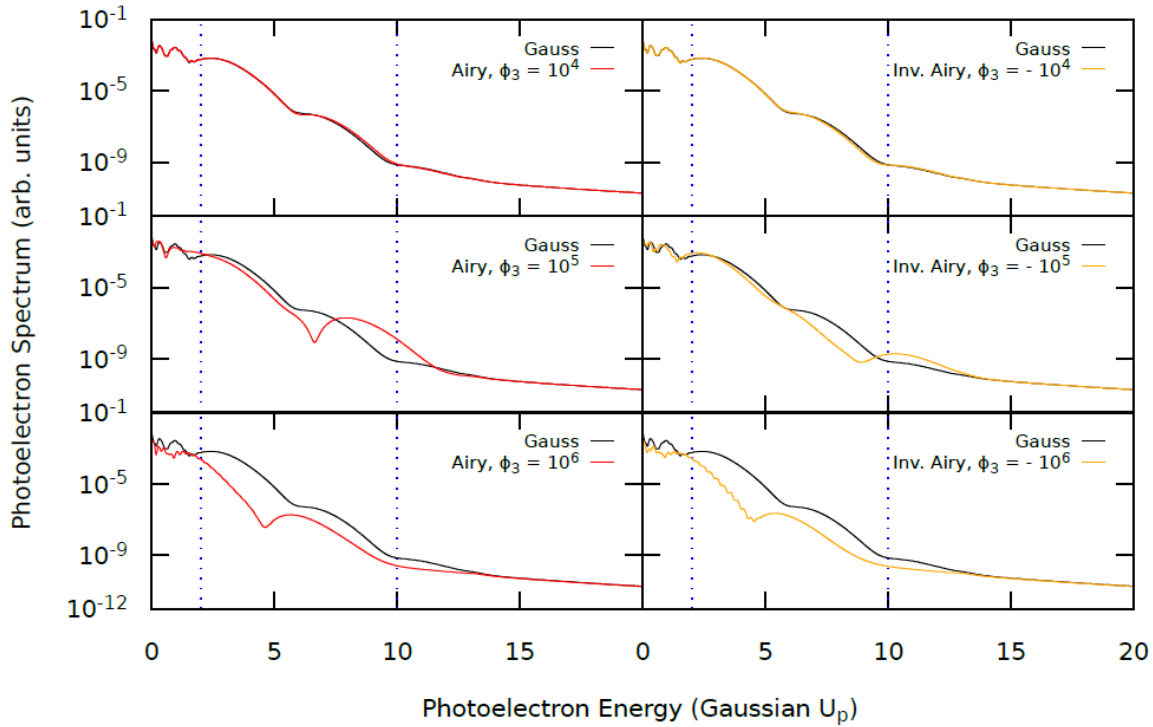

Figure 2 ATI photoelectron energy spectra for a hydrogen atom calculated with the 1D TDSE. The laser parameters are the same as Fig. 1. The black line is the spectrum for a Gaussian pulse (identical in all panels) and the red (orange) line is for a truncated Airy (inverted Airy) pulse. Third order phase values are listed in the figure. The blue dotted vertical lines represent the $2U_p$ and $10U_p$ values for the Gaussian pulse ($U_p = 5.96$ eV).

The results of Fig. 2 demonstrate that the spectral phase of the laser pulse alters the ATI energy spectrum. Additional evidence of the effects of the laser pulse's spectral phase on the photoelectron can be seen in the momentum density. Figure 3 shows the photoelectron momentum density as a function of momentum and laser pulse spectral phase. Data from the TDSE simulations are shown for $|\phi_3| \leq 10^5$, in which the pulse envelope most resembles that of a Gaussian and the ATI spectra for the different pulse types are most similar. A clear change in the momentum density is observable as $\phi_3$ changes, indicating that the spectral phase of the laser pulse can alter the photoelectron wave packet, despite nearly identical pulse envelope shapes. It is well-known that the carrier envelope phase alters the photoelectron momentum density, despite identical pulse envelopes. However, our results indicate that pulses with identical carrier envelope phases and nearly identical envelopes can still result in different momentum densities as a result of altering the spectral phase. Figure 3 shows a nearly linear change in the most likely momentum value as $\phi_3$ changes. As $|\phi_3|$ increases, the dominant momentum peak near zero splits into two peaks. For $\phi_3 > 0$, the dominant peak occurs at a positive momentum value, while for $\phi_3 < 0$, the dominant peak occurs at a negative momentum value.

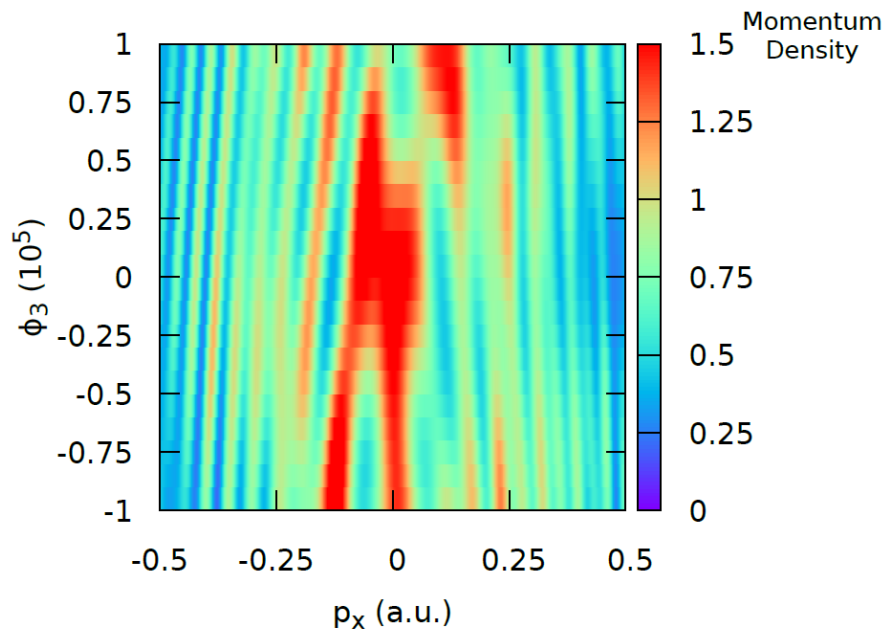

Figure 3 Photoelectron momentum density for ATI from hydrogen calculated using the time-evolved wave function for the 1D TDSE. The laser parameters are the same as Fig. 1. The momentum density (color) is plotted as a function of laser pulse spectral phase $\phi_3$ and photoelectron momentum $p_x$.

## B. Classical Simulations

By solving Newton's law for the motion of an electron in an oscillating electric field [43], the times of recollision ($t_{rec}$) as a function of ionization time ($t_{ion}$) can be found. The ionization time is defined as the time that the electron begins to move classically in the electric field, i.e. its 'birth' time. If the oscillating electric field drives the electron back to the parent ion, its position will return to the origin, and the time that this occurs is defined as the recollision time. It is possible that the electron returns to the origin at multiple instances as it is forced back and forth in the electric field, in which case there are multiple recollision times for a single ionization time.

Figure 4 shows a plot of the time difference between ionization and recollision $\Delta t$ (i.e. time after ionization) as a function of ionization time for the three pulse types. A single point on the graph indicates a $(t_{ion}, \Delta t)$ pair and a higher location of the point on the vertical axis indicates that the recollision occurs later after ionization. A vertical line drawn through the plot reveals that the number of $\Delta t$ values for a given $t_{ion}$ (i.e. vertical line crossings) equals the number of recollisions for a given ionization time. In all cases, there are more recollisions for electrons ionized earlier in the pulse than later in the pulse. This is expected given that the later an electron is ionized, the fewer cycles of the electric field it experiences, and thus the fewer recollisions.

Figure 4 shows that the different pulse shapes lead to different electron dynamics after ionization. For the inverted Airy pulse, there are fewer recollisions because the leading side lobes cause an increased electric field strength at earlier times. This enhanced electric field is able to drive the electron sufficiently far away from the origin that even the strong oscillations caused by the main lobe are too weak to bring the electron back to the parent ion. As $|\phi_3|$ increases, this effect is enhanced and fewer rescattering events are observed. These dynamics are expected to result in direct electrons coming from earlier ionization events and overall fewer rescattered electrons for the inverted Airy pulse, a result that is confirmed in the analysis of direct and rescattered electron kinetic energies shown below (Fig. 5).

In contrast, electrons ionized by the Airy pulse experience additional recollisions at later times compared to the Gauss and inverted Airy pulses. These additional recollisions are caused by the trailing side lobe of the Airy pulse, which enhances the electric field at later times relative to the Gaussian or inverted Airy pulses. This increased field strength at later times drives additional recollisions. Again, as $|\phi_3|$ increases, this effect is enhanced and more rescattering events are observed. In the case of the Airy pulse, the post-ionization dynamics are expected to result in fewer overall direct electrons and more rescattering at later ionization times, a result that is also confirmed in the kinetic energy spectra of Fig. 5.

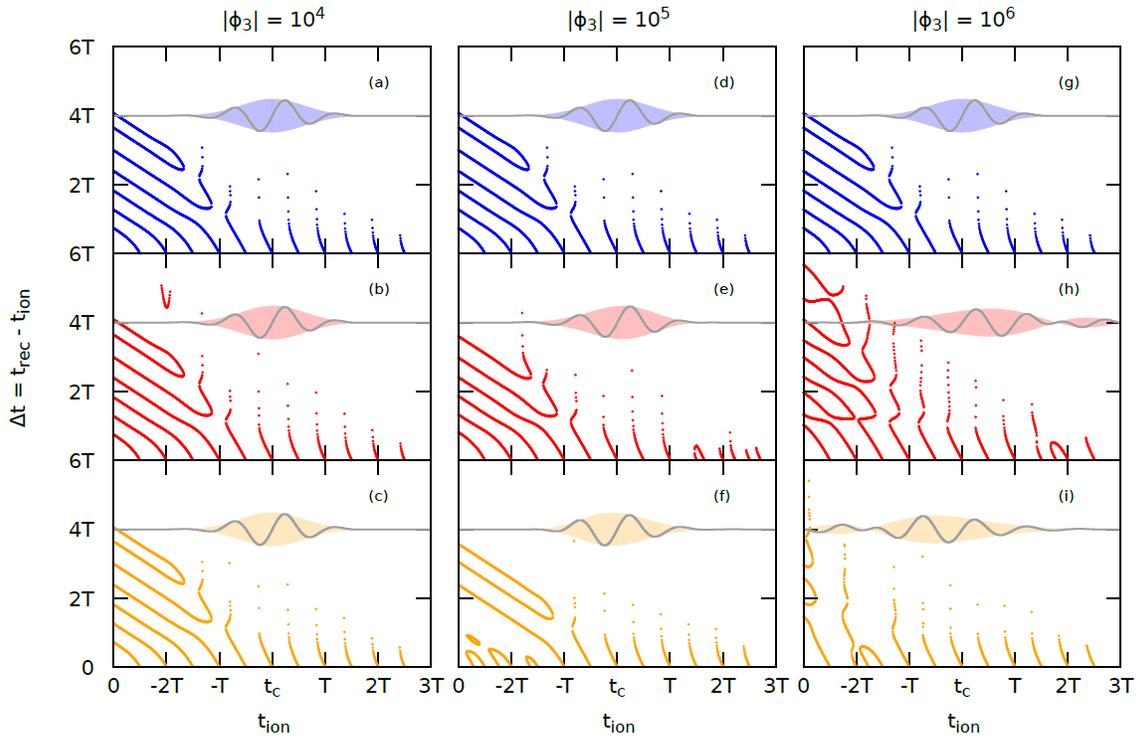

Figure 4 Time difference $\Delta t$ between ionization and recollision as a function of ionization time $t_{ion}$ calculated with the classical model. Each data point represents a recollision event. Times are in multiples of the electric field period ($T = 110$ a.u.) and $t_c$ is the temporal center of the pulse ($t_c = 330.5$ a.u.). The electric fields are shown in each panel along the horizontal line at 4T. The third order spectral phase is listed at the top of each column for the Airy and inverted Airy pulses, and the Gaussian plots in (a,d,g) are identical.

Further information about the dynamics of the electrons after ionization can be found by examining the kinetic energy of the ionized electrons at the detector. As discussed above, direct electrons have a lower maximum kinetic energy than rescattered electrons and therefore, the kinetic energy can be used to separate direct and rescattered electrons. Figure 5 shows the classical prediction for the kinetic energy of the direct and rescattered electrons as a function of the ionization time. The thin horizontal lines in each panel represent the $2U_p$ and $10U_p$ energy cutoffs for the different pulses (adjusted to account for changes to $U_p$ based on pulse type). Figure 4 shows that the maximum kinetic energies of the direct and rescattered electrons are approximately predicted by the $2U_p$ and $10U_p$ values, as long as the pondermotive energy is calculated using the maximum of the electric field strength of the particular pulse type used.

The laser pulse shape has a clear effect on the shape of the kinetic energy spectrum for the direct electrons, particularly for large $|\phi_3|$. The maximum in the direct electron kinetic energy spectrum coincides with the maximum of the pulse envelope and therefore, the direct electrons with the largest kinetic energy are produced at earlier times for the inverted Airy pulse and later times for the Airy pulse. The kinetic energy spectrum of direct electrons produced by Airy and inverted Airy pulses is also asymmetric with respect to the pulse's temporal center.

The peak value of the electric field envelope is shifted to later times for the Airy pulse and earlier times for the inverted Airy pulse. This asymmetry causes the electrons ionized by the Airy pulse after the temporal center to have greater kinetic energy than those ionized before, while the opposite is true for the inverted Airy pulse. The direct electron kinetic energy spectrum also shows the multiple peak structure of the pulse envelopes when $|\phi_3| = 10^6$. The secondary lobe of the Airy pulse results in the production of direct electrons at later ionization times that are not present for the Gaussian or inverted Airy pulse. Likewise, the inverted Airy pulse produces earlier direct electrons that are not present with the Gaussian or Airy pulses.

The laser pulse shape also affects the kinetic energy spectrum for the rescattered electrons, with most differences observed for the largest $|\phi_3|$ value. As was shown in Fig. 4, the Airy pulse produces electrons that undergo more rescatterings, than those produced by the inverted Airy and Gaussian pulses. The increased number of rescatterings by electrons ionized early from an Airy pulse leads to greater kinetic energy of these electrons compared to those ionized by a Gaussian pulse at the same time. Additionally, the trailing second lobe of the Airy pulse for $|\phi_3| = 10^6$ leads to rescattered electrons with non-zero kinetic energy produced at later times than those of a Gaussian pulse.

For the inverted Airy pulse, the emission of electrons that undergo rescattering occurs in two groupings, corresponding to the two lobes of the pulse envelope. The initial lobe is present for $t_{ion} < -2T$ and leads to electrons that rescatter with kinetic energy up to $10U_p$ (with $U_p$ calculated for the inverted airy electric field maximum). This is followed by a second lobe for $t_{ion} > -1.5T$, which leads to the second set of electrons that rescatter. The rescattered electrons ionized later by the inverted Airy pulse gain less kinetic energy than those ionized at the same times by the airy pulse, a direct effect of the pulse envelope asymmetry.

The lower energy cutoffs observed in the TDSE ATI spectrum in Fig. 2 are explained by the lower energy direct and rescattered electrons predicted by the classical model with reduced pondermotive energy. Lower energy direct electrons produced by Airy and inverted Airy pulses cause the Keldysh plateau cutoff to shift to lower energy compared to that of the Gaussian pulse. Likewise, the lower maximum kinetic energies of electrons generated by the Airy and inverted Airy pulses results in a shift to lower energy of the rescattering plateau relative to the Gaussian pulse.

While the three pulse types used here have identical spectral intensities for $|\phi_3| < 10^5$, their third order spectral phases and temporal envelopes differ. As a result of these changes, the dynamics of the ATI process are altered. For $|\phi_3| > 10^5$, the spectral intensity of the laser pulse no longer exactly matches that of a Gaussian pulse because we truncate the pulse at 6 cycles. For $|\phi_3| = 10^6$, rescattering dynamics are a result of the multiple peak structure of the Airy and inverted Airy pulses and reduced photoelectron kinetic energies result from decreased field strength of the sculpted pulses.

Recent studies have shown that sculpted photoelectrons in the form of Bessel electrons can be produced through the interaction of sculpted laser pulses with atoms [32,48,49], and investigations of the effect of these sculpted photoelectron wave packets on the HHG process have shown that they can alter the polarization and propagation directions of the emitted pulses [27,28]. Our results provide evidence that the photoelectron momentum and energy are altered through changes to the laser pulse spectral phase, as are the dynamics of ATI photoelectrons. It is therefore expected that future investigations of high-order processes, such as HHG, using Airy pulses will result in changes to the dynamics of these processes.

The ATI process can also be used to provide information about the target atom and parent ion during the recollision process [5], and comparison of ATI spectra for different sculpted laser pulses may be useful in providing more detailed information about the target. Sculpted electron wave packets have been shown to be useful probes of oriented targets, angular momentum-dependent effects, or multi-electron effects [50–57], and this provides an incentive to determine if similar properties can be leveraged for laser pulses sculpted through the adjustment of the spectral phase.

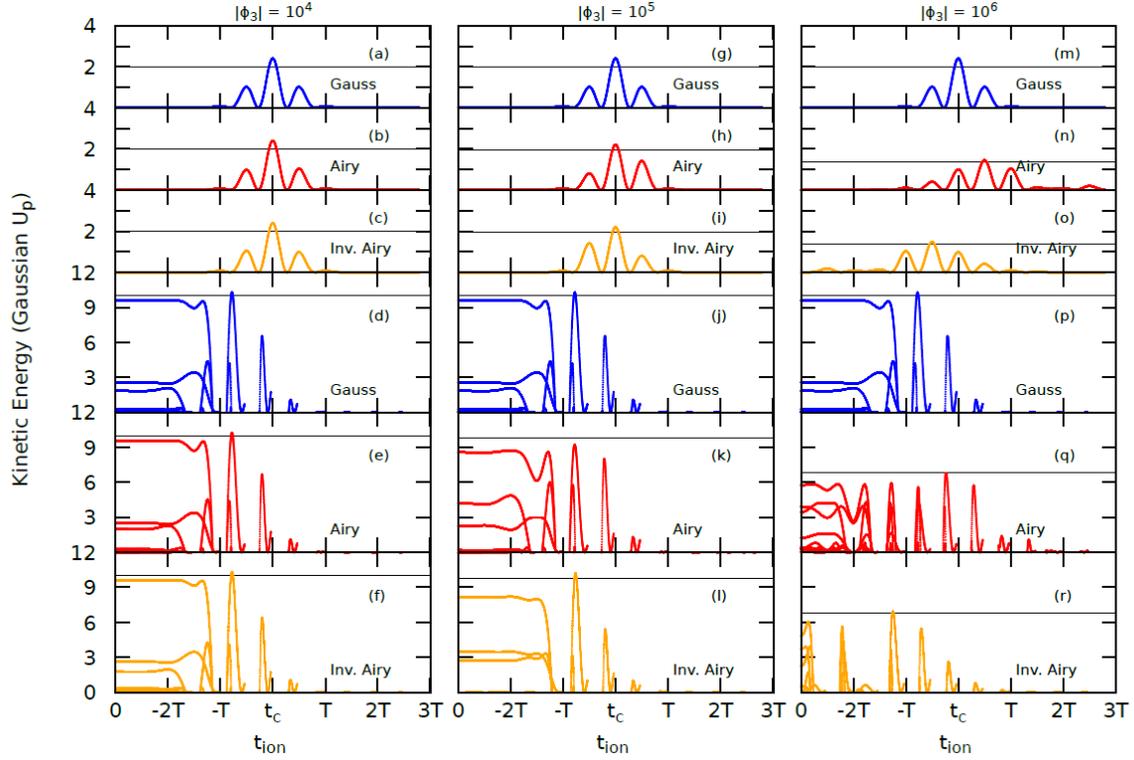

Figure 5 Kinetic energy spectra for direct (a-c, g-i, m-o) and rescattered (d-f, j-l, p-r) ATI electrons as a function of ionization time. Results were calculated with the classical model and $t_{ion}$ is given in units of the electric field period ($T = 110$ a.u.). The kinetic energies are shown in units of the Gaussian pulse ponderomotive energy ($U_p = 5.96$ eV). The thin horizontal lines in each panel represent the $2U_p$ and $10U_p$ cutoffs adjusted for the different pulse types. Relative values of $U_p$ for the pulses are listed in Table 2.

## C. Noble Gas Atoms

To test our model predictions beyond the hydrogen atom, we have used the TDSE to compute the ATI spectra for all noble gas atoms using Gaussian, Airy, and inverted Airy pulses. The same laser parameters as in Figs. 1-5 were used, and the model potentials were given by Eq. (2). Results are shown in Figs. 6-10. Because the classical equations of motion depend only on the electron in the presence of the laser field, they are independent of target atom and the same as those discussed in III.B. Thus, Figs. 4-5 apply to all target atoms.

In general, many of the conclusions of sections III.A do not change for different target atoms. For argon, krypton, and xenon, the Keldysh plateau cutoff shifts to lower energies as the magnitude of the third order spectral phase of the Airy and inverted Airy pulses increases. Because this feature was traced to the change in the pondermotive energy, it is a feature of the laser field and should be independent of target atom, as our results show.

The ATI spectra for helium and neon are quite different than those of hydrogen, argon, krypton, and xenon. For these atoms, no noticeable plateaus appear in the spectrum, but rather a mostly monotonic decay is observed with increasing energy. This feature has been observed previously in ATI spectra [1,47] and is attributable to the dependence of the photoelectron yield on the differential scattering cross section [1,2]. The energy dependence of the differential cross sections varies with target atom and this variation results in the observed changes in the ATI spectra.

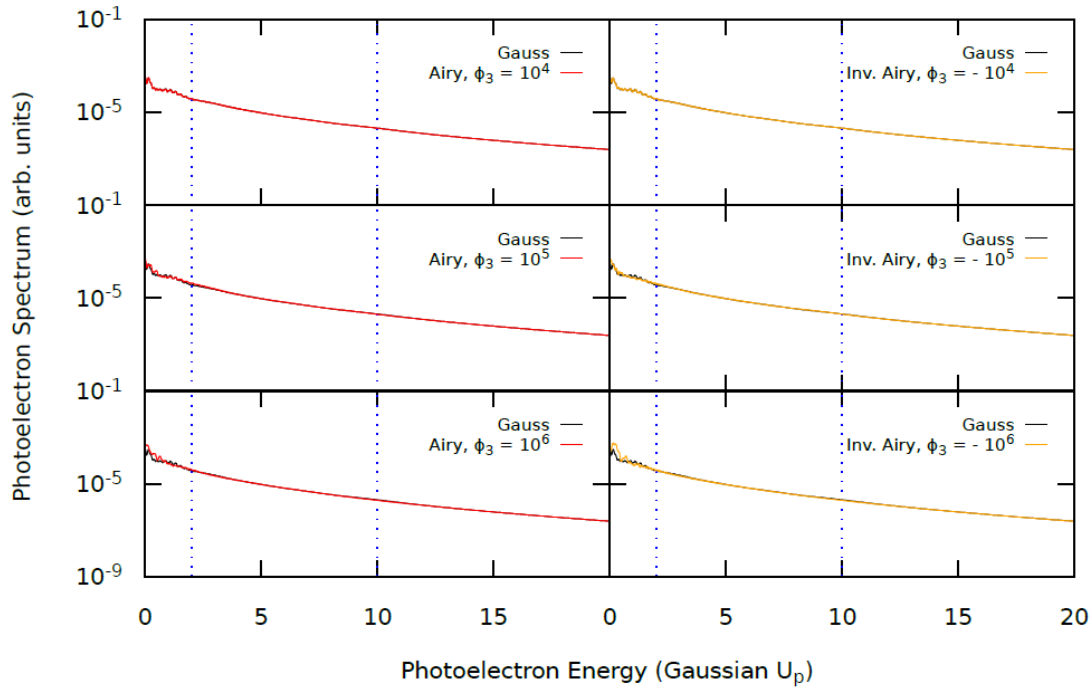

Figure 6 Same as Fig. 2 but for helium.

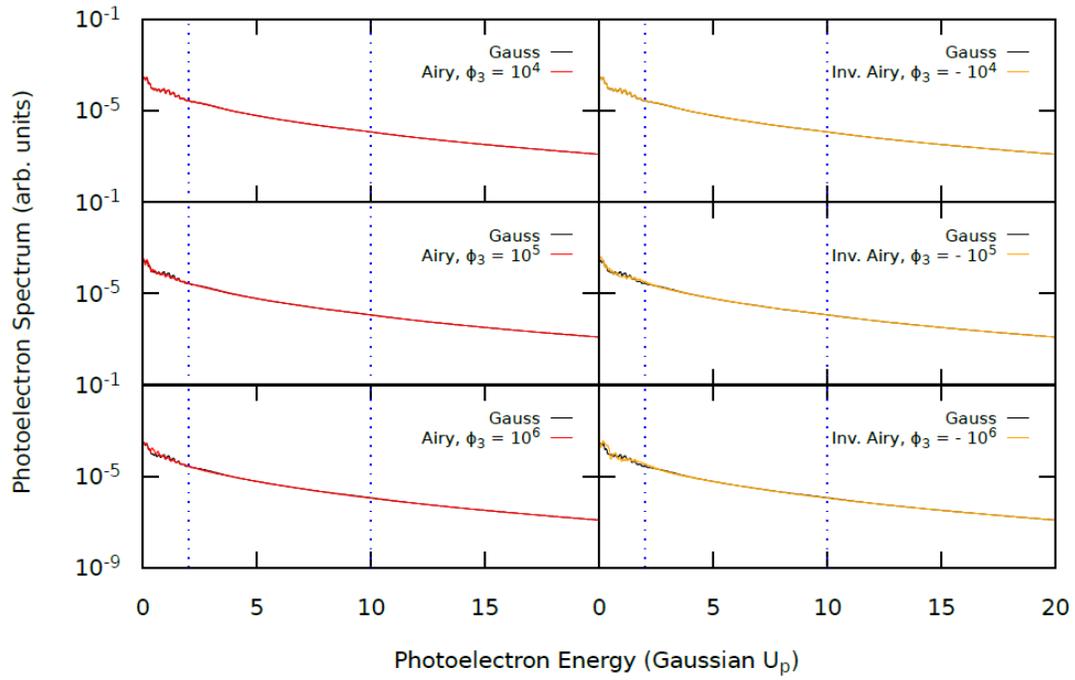

Figure 7 Same as Fig. 2 but for neon.

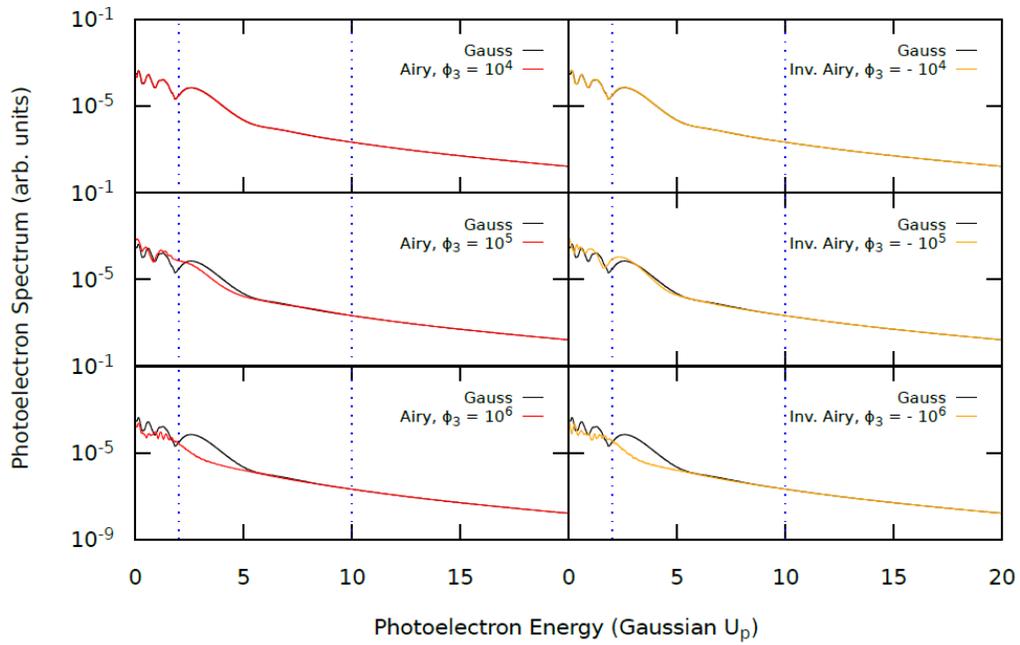

Figure 8 Same as Fig. 2 but for argon.

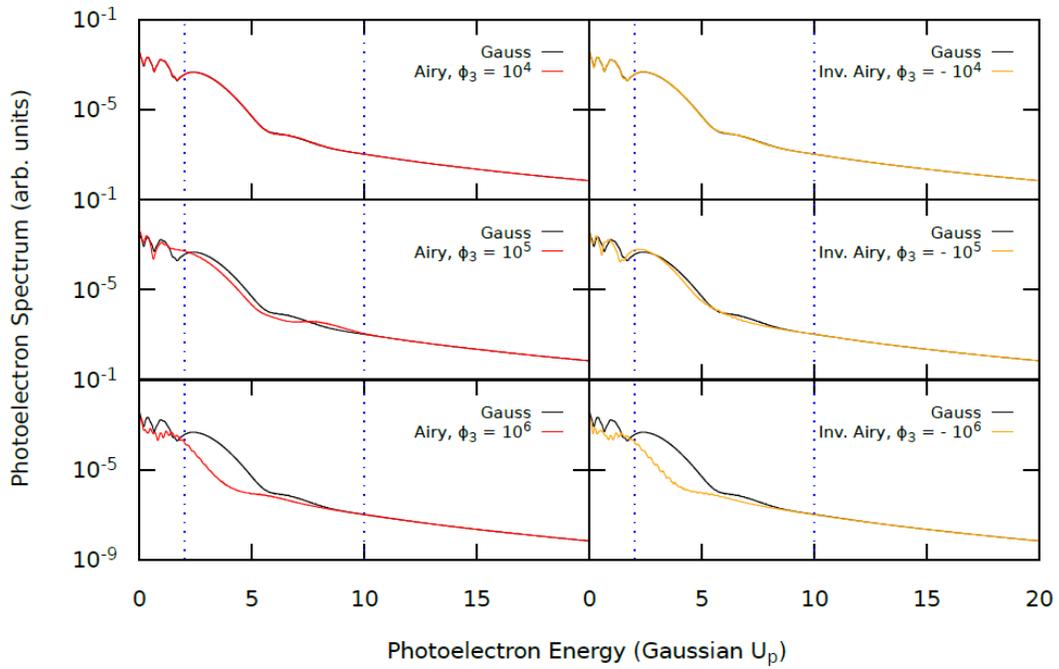

Figure 9 Same as Fig. 2 but for krypton.

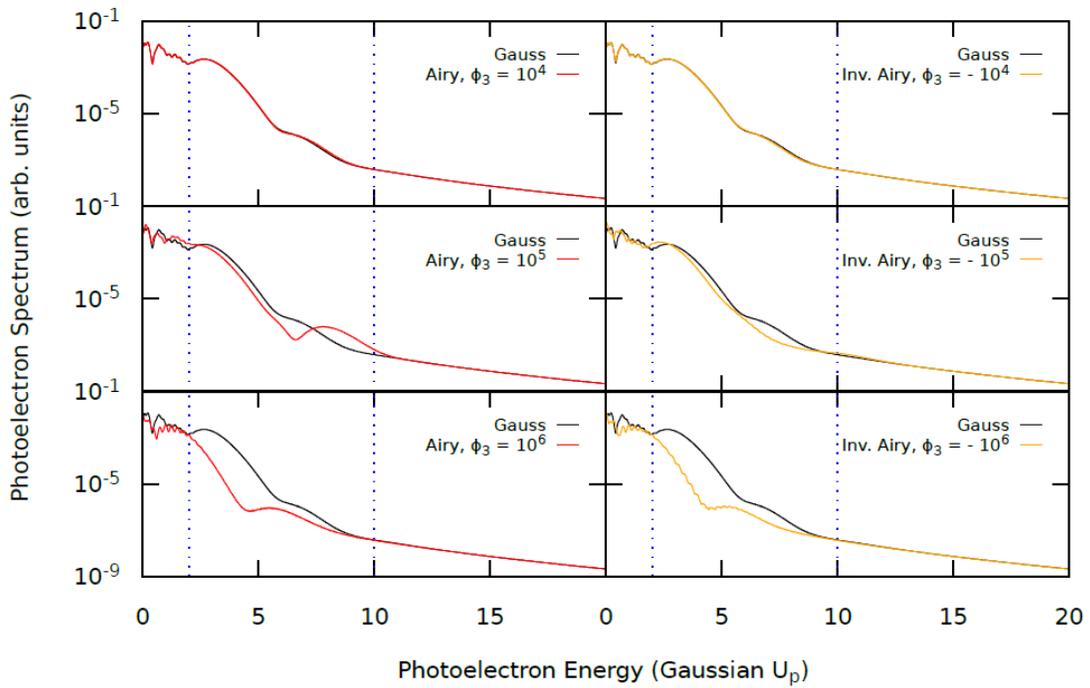

Figure 10 Same as Fig. 2 but for xenon.

**III. Summary**

By solving the TDSE, we have calculated above threshold ionization spectra for noble gas atoms in the presence of sculpted laser fields. Results were compared for electrons ionized by Gaussian, Airy, and inverted Airy pulses with identical power spectra and an in-depth analysis of the results was presented for a hydrogen target. These calculations showed that the Keldysh plateau cutoff and the high-order ATI plateau cutoff were shifted when either an Airy or inverted Airy pulse was used. For the Keldysh plateau, the shift of the cutoff energy to lower values was due to the reduced electric field strength, which reduced the ponderomotive energy. For the HATI plateau, the cutoff energy shifted to either higher or lower values for the Airy and inverted Airy pulses as a result of changes in rescattering dynamics.

Our TDSE results also showed that the momentum density of the photoelectron was altered with the use of the different laser pulses, despite the pulses having identical carrier envelope phases and nearly identical envelopes. We found an approximately linearly dependence of the photoelectron's most likely momentum value on the laser pulse's spectral phase, demonstrating that the laser pulse's spectral phase is imprinted on the photoelectron's momentum.

Using a classical model, we showed that the temporal pulse shape alters the post-ionization recollision dynamics. For the Airy pulse, there were more recollisions at later times due to the trailing side peaks of the pulse. In contrast, for the inverted Airy pulse, there were overall fewer recollisions due to the leading side lobes of the pulse causing ionization that drove the electron far from the origin. The classical model was also able to correctly predict the maximum kinetic energy of the direct and rescattered electrons when the change in ponderomotive energy due to pulse shape was considered.

The ATI spectra for targets other than hydrogen showed similar trends, with shifts in the energy plateaus observed for argon, krypton, and xenon. However, no significant changes were observed in the ATI spectra for different pulse types for helium and neon targets. The dependence of the ATI spectrum on target atom identity is well-documented and our results demonstrate that this does not change with the introduction of sculpted pulses.

Given that the ATI process is ubiquitous in ultrafast physics, the results presented here provide qualitative and quantitative insight into the effect of temporally sculpted laser pulses on the dynamics of the ATI process. These results lay the groundwork for applications of sculpted laser pulses in processes such as high order harmonic generation, non-sequential double ionization, streaking, RABBITT, and the attoclock. The presence of the third order spectral phase term in the complex electric field spectrum of the Airy pulse provides an additional control parameter for tuning the temporal field, while maintaining the spectral intensity.

**Appendix A**
**A. TDSE**

The 1D TDSE was solved using the Crank-Nicolson method [36]. The initial state wave function was evolved from $t = 0$ to $t = 661$ a.u. (the end of the laser pulse) with a time step of 0.01 a.u. The spatial grid spanned from $-3000$ a.u. to $3000$ a.u. with a step size of 0.033 a.u. Absorbing boundary conditions were employed [37] in the form of a mask function that prevented unphysical reflections at the boundaries

$$g(x) = \cos^{1/8}\left(\frac{\pi(|x|-(x_{max}-50.))}{2(50)}\right), \tag{A1}$$

where $x_{max}$ is the boundary edge (i.e. 3000 a.u.). At each time step, the wave function within 50 a.u. of the boundary (i.e. $x < -2950$ a.u. or $x > 2950$ a.u.) was multiplied by $g(x)$.

## B. Initial state wave function

The initial state atomic wave function was found by imaginary time propagation using an algorithm similar to that of the Crank-Nicolson method. The time step was 0.01 a.u. The spatial grid spanned from – 500 a.u. to 500 a.u. with a step size of 0.1 a.u. The initial guess of the wave function was [58]

$$\psi(x,0) = \left(1 + \sqrt{x^2+1}\right)e^{-\sqrt{x^2+1}}. \tag{A2}$$

## C. Window Method

The ATI energy spectrum was calculated using the window operator technique [38,39]. The energy bin width was $2\gamma = 0.02$ and the integer power for the window operator was $n = 2$.

## D. Classical Calculations

The classical calculations for finding the recollision times and kinetic energies of the direct and rescattered electron were calculated using a modified version of the ClassSTRONG program [43]. Simulations for the recollision times were run from $t = 0$ to a final time of $t = 661$ a.u. with 1000 time steps. For the kinetic energies, the simulations were run from $t = 0$ a.u. to $t = 661$ a.u. with 7000 time steps.

**Acknowledgements**
We gratefully acknowledge the support of the National Science Foundation under Grant No. PHY- 2207209.